# Defining AI Fatigue in Academic Contexts: Dimensions, Indicators, and a Stage-Based Model Using Grounded Theory

**John Paul P. Miranda*, Emmanuel B. Parreño and Jovita G. Rivera**
Pampanga State University
Pampanga, Philippines

**Abstract.** The integration of AI tools in academic settings has introduced a distinct form of strain that existing frameworks like technostress and digital fatigue have not yet fully addressed. This study develops a conceptual model and identifies the dimensions that define AI fatigue as a form of strain arising from sustained academic use of AI tools. Using grounded theory analysis of open-ended responses from 1,054 university students across three universities in the Philippines, the study examined the cognitive, motivational, emotional, physical, and attentional pressures students experienced during AI-supported academic work. Analysis produced five dimensions of AI fatigue, namely Cognitive Overload, Motivational Disengagement, Moral Unease, Physical Strain, and Attentional Drift, each consisting of two indicators grounded in participant accounts. The findings also yielded the AI Fatigue Model, a stage-based framework that explains how these pressures accumulate and reinforce one another across repeated AI interaction in academic tasks. These contributions establish a conceptual and exploratory foundation for AI fatigue as a distinct construct and provide a basis for future instrument validation, scale development, and cross-contextual inquiry in academic settings where AI now mediates student learning.

*Corresponding author: John Paul P. Miranda; *jppmiranda@pampangastateu.edu.ph*

## 1. Introduction

Students now use Artificial Intelligence (AI) tools in many academic tasks, and this shift exposes them to pressures that existing research has not yet defined or explained. Studies on technostress show that digital tasks can create overload and mental strain, with overuse of technology consistently associated with elevated stress among university students (Cazan et al., 2024; Nisafani et al., 2020). Research on digital fatigue similarly found that learners experienced eye strain, frustration, and reduced attention during extended digital work (Dhir et al., 2018; Şambel Aykutlu et al., 2024), and Saleem et al. (2024) observed lower learning quality among online learners who faced demanding technological requirements. These studies establish that digital environments can generate strain, yet they do not address the pressures that may arise when students interact with AI systems that produce rapid, lengthy, or inconsistent outputs.

Established frameworks for technology-related strain, including technostress models describing overload and reduced control (Kumar, 2024; Wang & Yao, 2025), digital fatigue models describing screen-driven exhaustion (Ibrahim, Khaled, et al., 2025; Zhang & Deng, 2025), and Zoom fatigue models describing sensory and cognitive strain during synchronous communication (Salim et al., 2022; Tolentino & Miranda, 2020; Usta Kara & Esroy, 2022), show that technology influences thinking, motivation, and well-being. These models, however, do not address AI-specific pressures that emerge during repeated prompting, reviewing, and verification of AI-generated content, offering only partial insight into the form of strain during sustained AI-supported academic work.

Technostress describes overload and reduced control during general technology use and does not address the cognitive demands of prompting, reviewing, and verifying AI-generated outputs (Kumar, 2024; Nisafani et al., 2020). Digital fatigue involves screen-driven exhaustion during extended digital work but does not account for the motivational and moral pressures that emerge when students repeatedly rely on AI to complete academic requirements (Ibrahim, Khaled, et al., 2025). Zoom fatigue captures sensory and cognitive strain during synchronous video communication and is limited to a single platform and a single mode of interaction (Moralista et al., 2022; Salim et al., 2022; Tolentino & Miranda, 2020; Usta Kara & Esroy, 2022).

AI fatigue, as examined in this study, differs from these constructs because it arises from a specific pattern of interaction: students repeatedly prompt an AI system, receive rapid and often lengthy outputs, verify those outputs against their own understanding, and cycle through this process across multiple academic tasks. The inconsistency of AI responses and the prevalence of hallucinated content require students to repeatedly check and compare outputs, adding cognitive demands that existing frameworks do not account for as reported by earlier studies (Kim et al., 2025; Sanz-Tejeda et al., 2026; Shi et al., 2025; Zhai et al., 2024). Whether this pattern produces a coherent and distinct form of strain is the central question this study addresses.

Evidence shows that AI tools influence cognition and behavior in ways that differ from traditional digital platforms. Zhai et al. (2024) found that reliance on AI dialogue systems was associated with weaker critical thinking and reduced independent reasoning across fourteen empirical studies, and Gerlich (2025) reported higher cognitive offloading and lower analytical persistence among participants who used AI tools for complex tasks. Tian and Zhang (2025) found that greater AI dependence was associated with lower critical thinking, with cognitive fatigue partially mediating this relationship, while Kim et al. (2025) noted that inconsistent AI responses increased the verification demands placed on students. Yang et al. (2025) further documented that the shift from instrumental AI use to cognitive dependence was associated with weakened independent thinking and reduced creative output. These pressures appear across separate studies but have not yet been described as a single coherent phenomenon.

Existing models and measurement instruments reveal further gaps because available tools assess general digital strain but do not measure the pressures linked with AI-supported academic tasks. For example, Daud (2025) found that hybrid technology adoption increased task pressure and switching demands among university students (Fabian et al., 2024; Ibrahim et al., 2025), yet the study did not address strain linked with AI-generated outputs. Current scales measure technostress, digital overload, or academic fatigue (Cao et al., 2025; Carmona-Halty et al., 2024; Morales-García et al., 2024; Tafesse et al., 2024; Yglesias-Alva et al., 2025), yet none of these tools assess strain from repeated interaction with AI systems. Research on generative AI use in academic writing has also documented emotional tensions around authorship and non-disclosure, showing that students experience discomfort when AI output quality exceeds their own perceived competence, which raises concerns about epistemic ownership and academic integrity (Qu et al., 2025).

No existing model explains how cognitive, motivational, emotional, physical, and attentional pressures may strengthen across sustained AI use, and no measurement instrument evaluates these pressures in a systematic way. Studies on technostress have begun to examine how strain develops through sequential psychological mechanisms, but these models focus on general technology demands rather than the specific interaction cycle produced by AI-generated outputs (Avcı, 2026). These gaps show that research has not yet conceptualized or measured the emerging form of strain that students may experience when they complete academic work with AI support.

The present study responds to this gap by examining student accounts to identify whether the pressures in AI-supported academic work form a coherent structure that differs from the patterns described in earlier research. Therefore, the study introduces the AI Fatigue Model as the first stage-based framework explaining how strain develops and progresses across AI-supported academic tasks, and identifies the dimensions and indicators of AI fatigue as an initial conceptual foundation for future instrument development. These contributions provide a basis for theoretical development, empirical validation, and instructional use in academic environments where AI now shapes student learning.

## 2. Methodology

### 2.1 Research Design

This study utilized a qualitative design guided by grounded theory. The design allowed the researchers to analyze student accounts of AI-related strain, determine whether the pressures formed a coherent structure, and develop a conceptual model grounded directly in participant accounts.

### 2.2 Research Locale, Population, and Sampling

The study took place in three universities in Pampanga, Philippines. The population included students who had used AI chatbots for academic activities. The study included three groups of participants: those who participated in face-to-face interviews, those who joined virtual interviews, and those who answered the same questions through an online survey. Teachers invited students to complete the survey through Google Forms, and campus administrators helped distribute the link within the institutions. Prior todata collection, the protocol in this study was subjected to ethics review and acquired the clearance in a university. The sample included 1,054 students (507 male and 547 female) with a mean age of 19.21 years old.

### 2.3 Research Instrument and Data Gathering Procedure

The first group of participants took part in face-to-face interview sessions conducted by an independent interviewer who was not part of the research team. Each session lasted approximately 15 to 20 minutes. Participants answered open-ended questions designed to explore their experiences with AI-related fatigue. Six main questions guided the discussion. The first question, "*Can you describe your experiences using AI chatbots such as ChatGPT or similar tools for your studies?*" gathered detailed accounts of how students interact with AI tools in their academic activities. The second question, "*What kinds of school tasks or activities do you usually use them for?*" identified the specific academic contexts where AI tools are most often used.

The third question, "*What do you usually feel or experience, whether physical, emotional, or mental, after using these AI tools frequently for academic work?*" explored the types of fatigue or strain that students experience after frequent AI use. The fourth question, "*Have you ever felt tired, overwhelmed, or unmotivated when using them?*" determined whether students experience emotional or motivational exhaustion related to AI use. The fifth question, "*In what ways has your frequent use of AI tools negatively affected your learning habits, motivation, or focus?*" examined possible behavioral or academic consequences of prolonged AI interaction. The final question, "*Do you notice any negative changes in how you study or approach academic tasks compared to before you started using AI tools?*" examined changes in students' learning approaches over time.

All questions were available in both English and Filipino. The Filipino version was verified by a Filipino language professor. Students were first presented with the English version but were informed that a Filipino translation was also available. For the survey distribution, the same set of questions was provided through Google Forms, allowing respondents to answer in either language. The forms were distributed within the schools through their deans, campus directors, and program chairpersons.

### 2.4 Data Analysis

The collected responses were analyzed using grounded theory following the three-stage coding procedure described by Corbin and Strauss (2008): open coding, axial coding, and selective coding. Two researchers independently read all responses before coding began to develop familiarity with the range of experiences participants described. In the open coding stage, they independently assigned descriptive labels to phrases and statements that reflected experiences of strain, difficulty, or discomfort during academic AI use, grounding each label directly in participant language. In the axial coding stage, related labels were grouped into clusters, with each cluster representing a specific and recurring form of pressure across participant accounts. These clusters formed the indicators.

In the selective coding stage, they identified broader categories that connected the indicators into coherent dimensions, retaining a dimension only when both researchers independently agreed that its indicators reflected a unified form of strain. When the two researchers produced different labels or groupings at any stage, a third researcher was consulted to reach a final decision. They also examined the sequence in which participants described these pressures across their AI use experiences, and this sequence informed the stage-based structure of the model that emerged from the analysis.

## 3. Results

The analysis of 1,054 participant responses produced five dimensions of AI fatigue through the three-stage grounded theory coding process described in the Methodology. This process produced ten indicators organized under five dimensions: Cognitive Overload, Motivational Disengagement, Moral Unease, Physical Strain, and Attentional Drift. The response counts in Table 1 reflect the number of participants whose accounts were coded under each dimension or indicator. Table 1 presents the five dimensions and their related indicators.

**Table 1: Core dimensions, indicators, and representative responses of AI fatigue**

| Dimensions (Total Responses) | Indicators | Representative Sample Responses (Total Responses) |
|---|---|---|
| Cognitive overload (310) | Output Overwhelm | "My mind is now having a hard time thinking or functioning." (65)<br>"My head feels heavy when AI gives long answers." (58)<br>"I get overwhelmed because I see too many options." (61) |
| | Verification Strain | "I feel mentally exhausted after checking so many outputs." (72)<br>"I need to read again and again because AI gives different answers." (41)<br>"It drains me because I must verify everything." (32) |
| Motivational disengagement (270) | Initiative Loss | "It makes me feel lazy to do my schoolwork." (74)<br>"I ask AI before I even try to think." (59)<br>"I find it hard to think without using AI." (51) |
| | Competence Doubt | "I lost confidence because I rely on it too much." (41)<br>"I feel stressed because I am not learning the same way." (44)<br>"I worry that I depend on it for everything." (33) |
| Moral unease (190) | Authorship Guilt | "It feels wrong because it is not my own effort." (29)<br>"I sometimes feel guilty because I rely on AI." (68)<br>"I feel uneasy because it feels like cheating." (37) |
| | Understanding Gap | "It bothers me that my output looks better than my actual understanding." (21)<br>"I feel stressed because I know I am not learning the same way." (44)<br>"I feel sad when I compare my work with AI's ideas." (28) |
| Physical strain (120) | Visual Fatigue | "My eyes hurt after reading many AI explanations." (47)<br>"My eyes get tired because I read too much on the screen." (23)<br>"My head feels painful because the information is not accurate." (32) |
| | Session Exhaustion | "I get physically drained after long sessions with AI." (24)<br>"I feel tired from sitting too long using AI." (11)<br>"I feel restless after checking AI too many times." (6) |

| Dimensions (Total Responses) | Indicators | Representative Sample Responses (Total Responses) |
|---|---|---|
| Attentional drift (125) | Pace Impatience | "I get impatient when I need to read my textbook after using AI." (19)<br>"AI gives answers too fast and I lose patience with slower materials." (14)<br>"I feel restless when switching back to normal reading." (11) |
| | Restless Switching | "I lose focus because I jump from one question to another." (27)<br>"I keep switching tabs because I want faster answers." (38)<br>"I get overwhelmed when AI gives too much information." (33) |

Cognitive overload was reflected in participant accounts of difficulty processing AI-generated outputs and performing repeated verification tasks. Participants described mental strain arising from the volume and density of AI responses, with statements such as "*My mind is now having a hard time thinking or functioning*" (n = 65), "*My head feels heavy when AI gives long answers*" (n = 58), and "*I get overwhelmed because I see too many options*" (n = 61) reflecting the output overwhelm indicator. Verification strain was equally prominent, as participants reported exhaustion from repeated checking, stating "*I feel mentally exhausted after checking so many outputs*" (n = 72) and "*I need to read again and again because AI gives different answers*" (n = 41). A total of 32 participants described this repeated verification as draining because they felt compelled to confirm every output before using it academically.

Motivational disengagement was reflected in participant accounts of reduced initiative and growing reliance on AI for academic tasks. The initiative loss indicator was evident in statements such as "*It makes me feel lazy to do my schoolwork*" (n = 74), "*I ask AI before I even try to think*" (n = 59), and "*I find it hard to think without using AI*" (n = 51), which described a pattern of bypassing independent effort in favor of immediate AI assistance. The competence doubt indicator was reflected in accounts of eroding academic confidence, with participants stating, "*I lost confidence because I rely on it too much*" (n = 41) and "*I feel stressed because I am not learning the same way*" (n = 44). A total of 33 participants also expressed concern about the extent to which they had come to depend on AI across multiple academic requirements.

Moral unease was reflected in participant accounts of tension surrounding authorship and the perceived gap between AI-generated output and actual understanding. The authorship guilt indicator was evident in statements such as "*It feels wrong because it is not my own effort*" (n = 29), "*I sometimes feel guilty because I rely on AI*" (n = 68), and "*I feel uneasy because it feels like cheating*" (n = 37), which described discomfort arising from the use of AI-generated content in academic submissions. The understanding gap indicator was reflected in accounts of distress over the disparity between output quality and personal competence, with participants stating "*It bothers me that my output looks better than my actual understanding*" (n = 21) and "*I feel stressed because I know I am not learning the same* *way*" (n = 44). A total of 28 participants also described feeling discouraged when they compared their own ideas with those produced by AI.

Physical strain was reflected in participant accounts of bodily discomfort arising from reading and verifying AI-generated outputs during academic work. The visual fatigue indicator was evident in statements such as "*My eyes hurt after reading many AI explanations*" (n = 47), "*My eyes get tired because I read too much on the screen*" (n = 23), and "*My head feels painful because the information is not accurate*" (n = 32), which described ocular and cranial discomfort associated with extended reading of AI-generated content. The session exhaustion indicator was reflected in accounts of physical depletion following prolonged AI use, with participants stating, "*I get physically drained after long sessions with AI*" (n = 24) and "*I feel tired from sitting too long using AI*" (n = 11). A total of six participants also described restlessness after repeated checking of AI responses.

Attentional drift was reflected in participant accounts of difficulty sustaining focus during and after AI-supported academic work. The pace impatience indicator was evident in statements such as "*I get impatient when I need to read my textbook after using AI*" (n = 19), "*AI gives answers too fast and I lose patience with slower materials*" (n = 14), and "*I feel restless when switching back to normal reading*" (n = 11), which described a reduced tolerance for the pace of traditional learning materials following exposure to the rapid response speed of AI tools. The restless switching indicator was reflected in accounts of fragmented attention and tab-switching behavior, with participants stating, "*I lose focus because I jump from one question to another*" (n = 27) and "*I keep switching tabs because I want faster answers*" (n = 38). A total of 33 participants also described feeling overwhelmed when AI returned large volumes of information.

The AI Fatigue Model reflects the progression of student experiences across repeated academic AI use. In the Initial Effort Reduction stage, students described how AI lowered early cognitive load and allowed quick progress on simple tasks, with statements such as "*I ask AI before I even try to think*" (n = 59) and "*I find it hard to think without using AI*" (n = 51) reflecting how AI became a first response to academic demands rather than a supplementary tool. As AI use expanded into more complex work such as generating explanations and organizing ideas, students' exposure to long and dense outputs increased, marking the Expansion of AI Roles stage.

In the Accumulation of Strain stage, students began experiencing pressures across the five dimensions, particularly cognitive overload and physical strain, as reflected in statements such as "*I feel mentally exhausted after checking so many outputs*" (n = 72) and "*I need to read again and again because AI gives different answers*" (n = 41). The Disruption of Self-Regulation stage followed, with students describing difficulty maintaining study routines and resisting AI reliance, as reflected in "*It makes me feel lazy to do my schoolwork*" (n = 74) and "*I lost confidence because I rely on it too much*" (n = 41). In the Fatigue Consolidation stage, strain became heavier and more persistent, with moral unease and attentional drift becoming more prominent alongside cognitive and motivational pressures. The Feedback Loop represents the return to earlier stages when AI use continues, reinforcing and intensifying all pressures across each subsequent cycle. Figure 1 shows the full structure of this model.

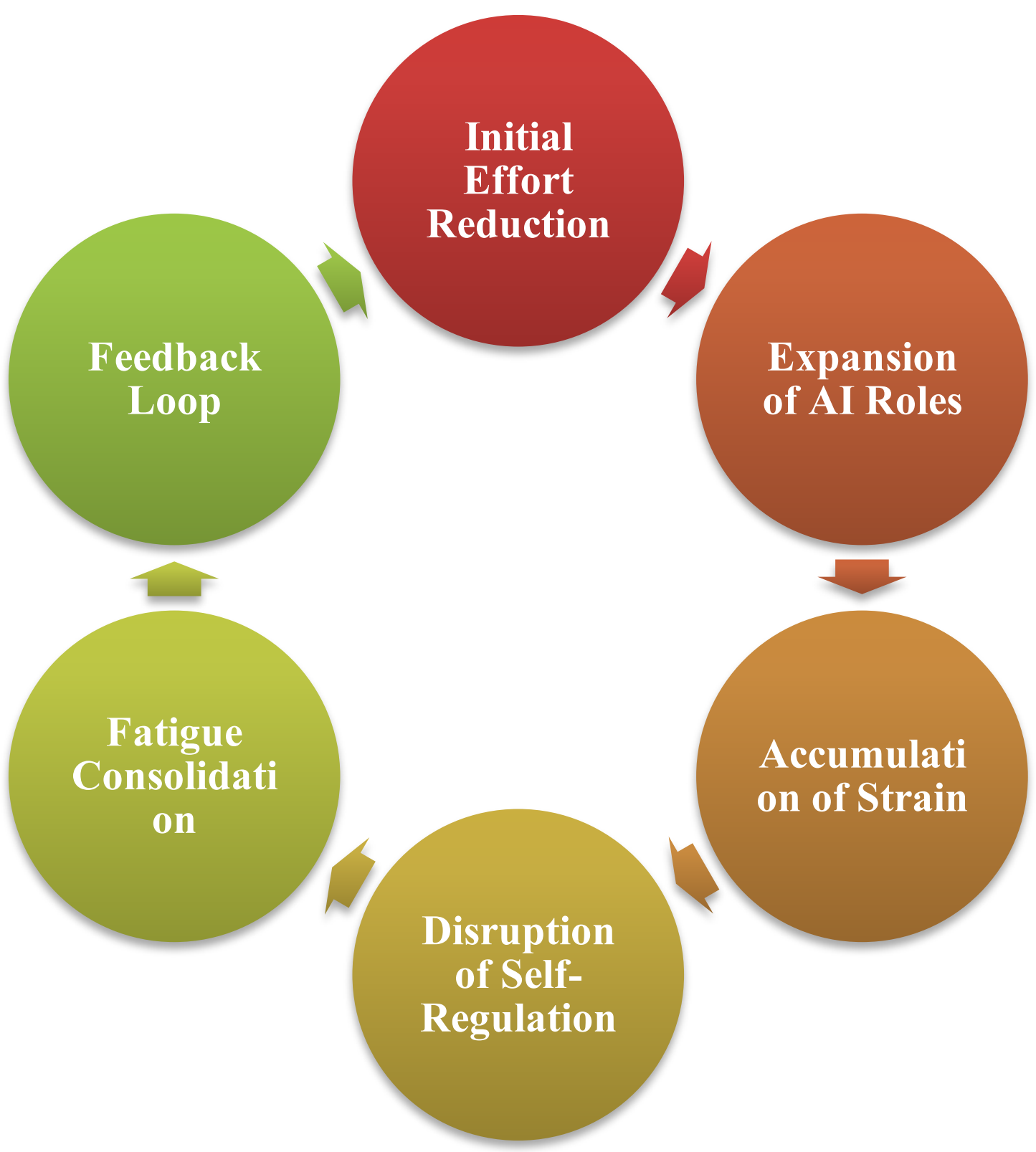

**Figure 1: Conceptual Model of AI Fatigue**

## 4. Discussion

This study identified five dimensions of AI fatigue that emerge from sustained academic use of AI tools among university students: Cognitive Overload, Motivational Disengagement, Moral Unease, Physical Strain, and Attentional Drift. Each dimension was grounded in participant accounts and supported by two indicators that captured specific and recurring forms of pressure during AI-supported academic work. Cognitive Overload was the most frequently reported dimension (n = 310), followed by Motivational Disengagement (n = 270), Moral Unease (n = 190), Attentional Drift (n = 125), and Physical Strain (n = 120).

These findings align with and extend earlier research on technology-related strain, cognitive offloading, and academic integrity in AI-supported learning environments (Gerlich, 2025; Manalese et al., 2025; Nastjuk et al., 2024; Qu et al., 2025; Zhai et al., 2024). The results also produced the AI Fatigue Model, a stage-based framework that explains how these pressures accumulate and reinforce one another across repeated AI interaction, distinguishing AI fatigue from technostress and digital fatigue as a coherent and progressively developing form of strain.

### 4.1 Cognitive Overload

Cognitive Overload was the most prominent dimension, with 310 participants reporting mental strain from processing AI-generated outputs and performing repeated verification tasks. The two indicators, Output Overwhelm and Verification Strain, reflected distinct but related sources of cognitive pressure, as participants described difficulty handling the volume and density of AI responses and exhaustion from repeatedly checking outputs for accuracy and consistency.

Tian and Zhang (2025) found that cognitive fatigue partially mediated the relationship between AI dependence and reduced critical thinking, suggesting that sustained engagement with AI-generated content accumulates mental effort over time. Triki and Turki (2025) similarly documented cognitive overload as a dimension of AI-related consumer fatigue, noting that output volume and complexity exceeded users' processing capacity during repeated interactions, while Shalu et al. (2025) reported that AI anxiety contributed to decision fatigue that shares structural similarities with the Output Overwhelm indicator identified in this study. These findings collectively suggest that the cognitive demands of academic AI use extend beyond learning the tool itself to the sustained effort of managing, evaluating, and verifying its content, positioning Cognitive Overload as a foundational pressure that may trigger the other dimensions of AI fatigue.

### 4.2 Motivational Disengagement

Motivational Disengagement was the second most reported dimension, with 270 participants describing reduced initiative and growing reliance on AI for academic tasks. The two indicators, Initiative Loss and Competence Doubt, captured a pattern in which students bypassed independent effort in favor of immediate AI assistance and experienced eroding confidence in their own academic capabilities, with many expressing concerns that their reliance had extended beyond individual tasks to broader academic functioning. Zhai et al. (2024) found that over-reliance on AI dialogue systems was associated with weaker critical thinking and reduced independent reasoning across fourteen empirical studies, and Yang et al. (2025) further documented that the shift from instrumental AI use to deeper cognitive dependence was associated with weakened independent thinking and reduced creative output.

Abbas et al. (2024) found that ChatGPT usage was associated with increased procrastination tendencies and memory loss, suggesting that sustained AI reliance may also impair the cognitive retention that underpins independent academic functioning, while Bai and Wang (2025) found that GAI interaction quality and output quality significantly influenced students' learning motivation and academic self-efficacy. These findings collectively suggest that Motivational Disengagement operates at both the behavioral and self-concept levels, as students who rely heavily on AI not only bypass independent effort but also begin to question their own learning and academic capability.

### 4.3 Moral Unease

Moral Unease was the third dimension identified in this study, with 190 participants reporting tension surrounding authorship and a perceived gap between AI-generated output quality and their actual understanding. The two indicators, Authorship Guilt and Understanding Gap, reflected ethical and epistemic discomfort that emerged during sustained AI use, as participants described guilt over submitting AI-generated content as their own work and distress over the disparity between what AI produced and what they genuinely understood. Cotton et al. (2024) documented significant tension around authorship and academic honesty in AI-supported settings, and Dong et al. (2025) similarly recorded emotional discomfort and learning burnout among students who misused generative AI, suggesting that moral tension about AI use can develop into a persistent form of strain.

Qu et al. (2025) showed that students experienced discomfort when AI output quality exceeded their own perceived competence, which directly corresponds to the Understanding Gap indicator, while Seran et al. (2025) described generative AI-induced cognitive dissonance where students held conflicting beliefs about the value and legitimacy of AI-assisted work. Saracini et al. (2025) and Klimova and Pikhart (2025) also noted emotional and ethical tensions in human-AI relationships in educational contexts. The Moral Unease dimension extends these findings by showing that ethical discomfort during AI use is not simply about rule violation but also about students' awareness of a growing gap between their AI-assisted outputs and their actual learning, adding a distinctive epistemic dimension that earlier frameworks have not captured.

**4.4 Physical Strain**

Physical Strain was the fourth dimension identified in this study, with 120 participants reporting bodily discomfort arising from reading and verifying AI-generated outputs during academic work. The two indicators, Visual Fatigue and Session Exhaustion, captured distinct but related forms of physical discomfort, as participants described ocular and cranial discomfort from extended reading of AI-generated content and physical depletion following prolonged AI use sessions. Devi and Singh (2023) documented the hazards of excessive screen time on physical health, including eye strain, headaches, and postural fatigue, which correspond directly to both indicators identified in this study, and Şambel Aykutlu et al. (2024) similarly reported that extended digital media use produced significant effects on digital eye strain among adolescent and university learners.

Klimova and Pikhart (2025) also noted physical well-being concerns among students who engaged heavily with AI tools in higher education, further supporting the presence of this dimension in AI-supported academic contexts. The Physical Strain dimension extends these findings by sitting in physical discomfort specifically within the AI interaction context, where the demands of reading lengthy and dense outputs and performing repeated verification tasks place sustained demands on the body that differ from general screen fatigue.

**4.5 Attentional Drift**

Attentional Drift was the fifth dimension identified in this study, with 125 participants reporting difficulty sustaining focus during and after AI-supported academic work. The two indicators, Pace Impatience and Restless Switching, captured how repeated exposure to the rapid response speed of AI tools reduced students' tolerance for the pace of traditional learning materials and produced fragmented attention, as participants described growing impatience when returning to textbooks after using AI and habitual tab-switching behavior driven by a preference for faster answers. Dhir et al. (2018) documented reduced attention and frustration during extended digital work, and Şambel Aykutlu et al. (2024) reported that heavy digital media use disrupted sustained attention among adolescent and university students, establishing that attentional disruption is a recognized consequence of extended digital engagement.

Tafesse et al. (2024) found that digital overload was associated with reduced student engagement and increased switching behavior, which corresponds directly to the Restless Switching indicator, while Ibrahim, Al Marar, et al. (2025) found that multiple educational technologies increased digital cognitive load, suggesting that AI interaction compounds the attentional demands already produced by other digital tools. The Pace Impatience indicator extends earlier research by showing that AI tools do not merely distract students but actively recalibrate their expectations about the pace of information delivery, as the near-instantaneous response speed of AI systems creates a contrast effect that makes traditional reading and reasoning feel disproportionately slow, gradually eroding students' capacity for sustained independent study.

**4.6 AI Fatigue Conceptual Model**

The AI Fatigue Model proposes a stage-based explanation of how the five dimensions develop and reinforce one another across repeated academic AI use through six progressively intensifying stages, as described in the Results section. This structure distinguishes AI fatigue from technostress and digital fatigue, which describe strain as a general response to technology demands, whereas Avcı (2026) noted that existing technostress models focus on general technology demands rather than the specific interaction cycle produced by AI-generated outputs, and Nastjuk et al. (2024) similarly found that technostress frameworks did not account for the motivational, emotional, and attentional dimensions that emerge specifically from AI interaction (Miranda et al., 2025).

The stage-based structure is supported by participant accounts in this study: the Initial Effort Reduction and Expansion of AI Roles stages are consistent with findings from Gerlich (2025) and Yang et al. (2025), who documented how AI tools initially reduce cognitive effort before encouraging expanded and increasingly dependent use across more complex academic tasks. The Accumulation of Strain stage is consistent with Tian and Zhang (2025) and Shalu et al. (2025), who documented converging cognitive and decision fatigue as AI interaction intensified, while the Disruption of Self-Regulation stage aligns with the motivational depletion and competence doubt described by Zhai et al. (2024) and Gerlich (2025). The Fatigue Consolidation stage and Feedback Loop are supported by Cotton et al. (2024), Qu et al. (2025), and Tafesse et al. (2024), whose findings on ethical tensions, epistemic discomfort, and attentional disruption reflect the pressures that become most prominent in the later stages of repeated AI use, where continued engagement perpetuates and intensifies strain across all dimensions rather than allowing recovery.

The findings of this study are situated within a higher education context where AI adoption notably outpaces institutional readiness. An ASEAN Foundation report found that most students in the Philippines have used generative AI tools for educational purposes, yet only half of schools in the region provide adequate training and clear usage guidelines, and fewer than half of educators express confidence in institutional AI policies and governance frameworks (Barro II, 2026; Manalese et al., 2025; Quiambao, 2026). This gap creates conditions where students use AI tools extensively without structured guidance on managing the pressures that sustained use produces, making the dimensions identified in this study, particularly Motivational Disengagement and Moral Unease, especially relevant in this context. These conditions are not unique to the Philippines, as most students across ASEAN rely on AI for information search and writing assistance while institutional readiness remains consistently low across the region (Rosales, 2026), suggesting that the AI Fatigue Model may have transferability to other ASEAN countries where high AI adoption similarly outpaces institutional preparedness.

The study also carries practical implications for researchers, educators, and institutions. Researchers can use the five dimensions and the stage-based model as a conceptual basis for item generation in scale development and for examining how strain accumulates across different levels of AI use intensity. Educators can use the model to identify points in the academic task cycle where cognitive or motivational pressure is likely to increase and adjust task design, pacing, and scaffolding accordingly, such as building structured reflection activities that prompt students to evaluate their own reasoning independently of AI output. Institutions can use the model to design AI literacy programs that address the psychological and motivational pressures of sustained AI use, supporting more regulated and intentional academic AI engagement.

## 5. Conclusion

This study addressed a gap in existing technology-related strain frameworks by identifying five dimensions of AI fatigue, namely Cognitive Overload, Motivational Disengagement, Moral Unease, Physical Strain, and Attentional Drift, through grounded theory analysis of responses from 1,054 students across three universities in the Philippines. Unlike technostress, digital fatigue, and Zoom fatigue frameworks, which describe strain as a general response to technology demands, the AI Fatigue Model explains how these pressures accumulate and reinforce one another through a six-stage progression, constituting the first conceptual framework specific to AI fatigue in academic contexts. The five dimensions and the stage-based model provide a foundation for instrument development, scale validation, and cross-contextual empirical inquiry, and the model can guide educators and institutions in designing pedagogical interventions and AI literacy programs that address the psychological and motivational pressures produced by sustained academic AI use.

The findings are exploratory, as the sample was drawn from three universities in one academic region and the analysis relied on self-reported data, and the model requires empirical validation before broader application. Future research should validate the construct through confirmatory factor analysis, examine how the model manifests across varied cultural and institutional contexts, test its applicability with domain-specific AI tools, and determine whether the five dimensions and stage-based progression generalize to other ASEAN countries where AI adoption similarly outpaces institutional preparedness.

**Conflict of Interest**
The author declared no conflicts of interest related to the research, authorship, or publication of this article.

**Acknowledgments**
This study was supported by Pampanga State University.

## 6. References

Abbas, M., Jam, F. A., & Khan, T. I. (2024). Is it harmful or helpful? Examining the causes and consequences of generative AI usage among university students. *International Journal of Educational Technology in Higher Education*, *21*(1), 10. https://doi.org/10.1186/s41239-024-00444-7

Avcı, M. (2026). The double-edged sword of technology: Investigating technostress and techno-eustress in academic burnout through digital literacy, internet self-efficacy, and cognitive flexibility. *Computers & Education*, *245*, 105540. https://doi.org/10.1016/j.compedu.2025.105540

Bai, Y., & Wang, S. (2025). Impact of generative AI interaction and output quality on university students' learning outcomes: a technology-mediated and motivation-driven approach. *Scientific Reports*, *15*(1), 24054. https://doi.org/10.1038/s41598-025-08697-6

Barro II, D. (2026, February 10). Filipino students embrace AI, but schools lag in readiness-ASEAN Foundation. *Manila Bulletin*. https://mb.com.ph/2026/02/10/filipino-students-embrace-ai-but-schools-lag-in-readinessasean-foundation

Cao, S., Ali, S., Yawar, R. B., Saif, N., Goh, G. G. G., Khan, F., & Hussain, M. (2025). Constructing and validating a scale for technostress and employee behavior: evidence from business schools in a developing country. *BMC Psychology*, *13*(1), 812. https://doi.org/10.1186/s40359-025-03152-7

Carmona-Halty, M., Alarcón-Castillo, K., Semir-González, C., Sepúlveda-Páez, G., & Schaufeli, W. B. (2024). Burnout Assessment Tool for Students (BAT-S): evidence of validity in a Chilean sample of undergraduate university students. *Frontiers in Psychology*, *15*, 1434412. https://doi.org/10.3389/fpsyg.2024.1434412

Cazan, A.-M., David, L. T., Truța, C., Maican, C. I., Henter, R., Năstasă, L. E., Nummela, N., Vesterinen, O., Rosnes, A. M., Tungland, T., Gudevold, E., Digernes, M., Unz, D., Witter, S., & Pavalache-Ilie, M. (2024). Technostress and time spent online. A cross-cultural comparison for teachers and students. *Frontiers in Psychology*, *15*, 1377200. https://doi.org/10.3389/fpsyg.2024.1377200

Corbin, J., & Strauss, A. (2008). *Basics of Qualitative Research: Techniques and Procedures for Developing Grounded Theory* (3rd ed.). SAGE Publications, Inc. https://doi.org/10.4135/9781452230153

Cotton, D. R. E., Cotton, P. A., & Shipway, J. R. (2024). Chatting and cheating: Ensuring academic integrity in the era of ChatGPT. *Innovations in Education and Teaching International*, *61*(2), 228–239. https://doi.org/10.1080/14703297.2023.2190148

Daud, N. M. (2025). From innovation to stress: analyzing hybrid technology adoption and its role in technostress among students. *International Journal of Educational Technology in Higher Education*, *22*(1), 31. https://doi.org/10.1186/s41239-025-00529-x

Devi, K. A., & Singh, S. K. (2023). The hazards of excessive screen time: Impacts on physical health, mental health, and overall well-being. *Journal of Education and Health Promotion*, *12*, 413. https://doi.org/10.4103/jehp.jehp_447_23

Dhir, A., Yossatorn, Y., Kaur, P., & Chen, S. (2018). Online social media fatigue and psychological wellbeing—A study of compulsive use, fear of missing out, fatigue, anxiety and depression. *International Journal of Information Management*, *40*, 141–152. https://doi.org/10.1016/j.ijinfomgt.2018.01.012

Dong, X., Wang, Z., & Han, S. (2025). Mitigating Learning Burnout Caused by Generative Artificial Intelligence Misuse in Higher Education: A Case Study in Programming Language Teaching. In *Informatics* (Vol. 12, Issue 2). https://doi.org/10.3390/informatics12020051

Fabian, K., Smith, S., & Taylor-Smith, E. (2024). Being in Two Places at the Same Time: a Future for Hybrid Learning Based on Student Preferences. *TechTrends*, *68*(4), 693–704. https://doi.org/10.1007/s11528-024-00974-x

Gerlich, M. (2025). AI Tools in Society: Impacts on Cognitive Offloading and the Future of Critical Thinking. In *Societies* (Vol. 15, Issue 1). https://doi.org/10.3390/soc15010006

Ibrahim, R. K., Al Marar, Y. A., Salman, M., Jehad, S., Hamza, M. G., Abouelnasr, A. S., Abdelaliem, S. M. F., Alahmedi, S. H., & Hendy, A. (2025). Impact of multiple educational technologies on well-being: the mediating role of digital cognitive load. *BMC Nursing*, *24*(1), 1028. https://doi.org/10.1186/s12912-025-03655-z

Ibrahim, R. K., Khaled, M., Almansoori, M., Almazrouei, M., Ashraf, A., Alahmedi, S. H., & Hendy, A. (2025). Screen time and stress: understanding how digital burnout influences health among nursing students. *BMC Nursing*, *24*(1), 990. https://doi.org/10.1186/s12912-025-03621-9

Kim, J., Yu, S., Detrick, R., & Li, N. (2025). Exploring students' perspectives on Generative AI-assisted academic writing. *Education and Information Technologies*, *30*(1), 1265–1300. https://doi.org/10.1007/s10639-024-12878-7

Klimova, B., & Pikhart, M. (2025). Exploring the effects of artificial intelligence on student and academic well-being in higher education: a mini-review. *Frontiers in Psychology*, *16*, 1498132. https://doi.org/10.3389/fpsyg.2025.1498132

Kumar, P. S. (2024). TECHNOSTRESS: A comprehensive literature review on dimensions, impacts, and management strategies. *Computers in Human Behavior Reports*, *16*, 100475. https://doi.org/10.1016/j.chbr.2024.100475

Manalese, R. P., Castro, M. A. A., Simpao, L. S., Liwanag, I. G., Sese, A. V. A., Arcilla, R. T., Garcia, E. D., & Miranda, J. P. P. (2025). Exploratory Factor Analysis of AI Chatbot Use in Supporting University Students' Academic Mental Health. *2025 1st International Conference on Emerging Trends in Information Systems and Informatics (ICETISI)*, 1–5. https://doi.org/10.1109/ICETISI67983.2025.11406053

Miranda, J. P. P., Nacianceno, M. C. B., Manalese, R. P., Liwanag, I. G., Simpao, L. S., Peñaflor, M. A. F., Valdez, V. V. G., & Acoba, M. T. (2025). Dimensions of AI-Induced Fatigue Among Filipino Students: An Exploratory Study. *2025 IEEE 23rd Student Conference on Research and Development (SCOReD)*, 1–4. https://doi.org/10.1109/SCOReD68498.2025.11399092

Morales-García, W. C., Sairitupa-Sanchez, L. Z., Morales-García, S. B., & Morales-García, M. (2024). Development and validation of a scale for dependence on artificial intelligence in university students. *Frontiers in Education*, *9*, 1323898. https://doi.org/10.3389/feduc.2024.1323898

Moralista, R., Oducado, R. M., Robles, B. R., & Rosano, D. (2022). Determinants of Zoom Fatigue Among Graduate Students of Teacher Education Program. *International*

*Journal of Emerging Technologies in Learning (IJET)*, *17*(13), 176–185. https://doi.org/10.3991/ijet.v17i13.31511

Nastjuk, I., Trang, S., Grummeck-Braamt, J.-V., Adam, M. T. P., & Tarafdar, M. (2024). Integrating and synthesising technostress research: a meta-analysis on technostress creators, outcomes, and IS usage contexts. *European Journal of Information Systems*, *33*(3), 361–382. https://doi.org/10.1080/0960085X.2022.2154712

Nisafani, A. S., Kiely, G., & Mahony, C. (2020). Workers' technostress: a review of its causes, strains, inhibitors, and impacts. *Journal of Decision Systems*, *29*(sup1), 243–258. https://doi.org/10.1080/12460125.2020.1796286

Qu, Y., Loo, H. E., & Wang, J. (2025). Generative artificial intelligence in higher education: Emotional tensions and ethical declaration. *British Journal of Educational Technology*, *n/a*(n/a). https://doi.org/10.1111/bjet.70029

Quiambao, J. (2026, February 11). Asean study shows high AI use among Pinoy students, flags readiness gaps. *NewsBytes.PH*. https://newsbytes.ph/2026/02/11/asean-study-shows-high-ai-use-among-pinoy-students-flags-readiness-gaps

Rosales, E. F. (2026, February 10). 3 in 4 Pinoy students use AI writing tools – poll. *Philstar Global*. https://www.philstar.com/headlines/2026/02/10/2506971/3-4-pinoy-students-use-ai-writing-tools-poll

Saleem, F., Chikhaoui, E., & Malik, M. I. (2024). Technostress in students and quality of online learning: role of instructor and university support. *Frontiers in Education*, *9*, 1309642. https://doi.org/10.3389/feduc.2024.1309642

Salim, J., Tandy, S., Arnindita, J. N., Wibisono, J. J., Haryanto, M. R., & Wibisono, M. G. (2022). Zoom fatigue and its risk factors in online learning during the COVID-19 pandemic. *Medical Journal of Indonesia*, *31*(1), 13–19. https://doi.org/10.13181/mji.oa.225703

Şambel Aykutlu, M., Aykutlu, H. C., Özveren, M., & Garip, R. (2024). Digital media use and its effects on digital eye strain and sleep quality in adolescents: A new emerging epidemic? *PloS One*, *19*(12), e0314390. https://doi.org/10.1371/journal.pone.0314390

Sanz-Tejeda, A., Domínguez-Oller, J. C., Baldaquí-Escandell, J. M., Gómez-Díaz, R., & García-Rodríguez, A. (2026). The impact of generative AI on academic reading and writing: a synthesis of recent evidence (2023–2025). *Frontiers in Education*, *10*, 1711718. https://doi.org/10.3389/feduc.2025.1711718

Saracini, C., Cornejo-Plaza, M. I., & Cippitani, R. (2025). Techno-emotional projection in human-GenAI relationships: a psychological and ethical conceptual perspective. *Frontiers in Psychology*, *16*, 1662206. https://doi.org/10.3389/fpsyg.2025.1662206

Seran, C. E., Tan, M. J. T., Abdul Karim, H., & AlDahoul, N. (2025). A conceptual exploration of generative AI-induced cognitive dissonance and its emergence in university-level academic writing. *Frontiers in Artificial Intelligence*, *8*, 1573368. https://www.frontiersin.org/journals/artificial-intelligence/articles/10.3389/frai.2025.1573368

Shalu, Verma, N., Dev, K., Bhardwaj, A. B., & Kumar, K. (2025). The Cognitive Cost of AI: How AI Anxiety and Attitudes Influence Decision Fatigue in Daily Technology Use. *Annals of Neurosciences*, 09727531251359872. https://doi.org/10.1177/09727531251359872

Shi, J., Liu, W., & Hu, K. (2025). Exploring How AI Literacy and Self-Regulated Learning Relate to Student Writing Performance and Well-Being in Generative AI-Supported Higher Education. *Behavioral Sciences (Basel, Switzerland)*, *15*(5), 705. https://doi.org/10.3390/bs15050705

Tafesse, W., Aguilar, M. P., Sayed, S., & Tariq, U. (2024). Digital Overload, Coping Mechanisms, and Student Engagement: An Empirical Investigation Based on the S-O-R Framework. *Sage Open*, *14*(1), 21582440241236090.

https://doi.org/10.1177/21582440241236087
Tian, J., & Zhang, R. (2025). Learners' AI dependence and critical thinking: The psychological mechanism of fatigue and the social buffering role of AI literacy. *Acta Psychologica*, *260*, 105725. https://doi.org/10.1016/j.actpsy.2025.105725
Tolentino, J. C. G., & Miranda, J. P. P. (2020). Self-reported zoom exhaustion and fatigue levels among physical education teacher education students in a state university in the Philippines. *Journal on Efficiency and Responsibility in Education and Science*, *17*(3), 205–213. https://doi.org/10.7160/eriesj.2024.170303
Triki, M., & Turki, A. M. (2025). Between Cognitive Overload and Dehumanization: Exploring the Dimensions of Consumer Fatigue with Artificial Intelligence. *International Journal of Research and Innovation in Social Science*, *9*(10), 7551–7564. https://doi.org/10.47772/IJRISS.2025.910000615
Usta Kara, I., & Esroy, E. G. (2022). A New Exhaustion Emerged with COVID-19 and Digitalization: A Qualitative Study on Zoom Fatigue. *OPUS Toplum Araştırmaları Dergisi*, *19*(46), 365–379. https://doi.org/10.26466/opusjsr.1069072
Wang, Q., & Yao, N. (2025). Understanding the impact of technology usage at work on academics' psychological well-being: a perspective of technostress. *BMC Psychology*, *13*(1), 130. https://doi.org/10.1186/s40359-025-02461-1
Yang, Z., Deng, H., & Jiang, N. (2025). The impact mechanism of artificial intelligence dependence on college students' innovation capability: an empirical study from China. *Frontiers in Psychology*, *16*, 1732837. https://doi.org/10.3389/fpsyg.2025.1732837
Yglesias-Alva, L. A., Estrada-Alva, L. A., Lizarzaburu-Montero, L. M., Miranda-Troncoso, A. E., Aguilar-Armas, H. M., & Vera-Calmet, V. G. (2025). Digital Fatigue in Peruvian University Students: Design and Validation of a Multidimensional Questionnaire. *Journal of Educational and Social Research*, *15*(3), 431. https://doi.org/10.36941/jesr-2025-0109
Zhai, C., Wibowo, S., & Li, L. D. (2024). The effects of over-reliance on AI dialogue systems on students' cognitive abilities: a systematic review. *Smart Learning Environments*, *11*(1), 28. https://doi.org/10.1186/s40561-024-00316-7
Zhang, X., & Deng, G. (2025). Protecting work engagement from digital fatigue: the contingent roles of leadership style and network ties. *Frontiers in Psychology*, *16*, 1645057. https://doi.org/10.3389/fpsyg.2025.1645057